\documentclass{mn2e}
\usepackage[]{epsfig}
\title{Magnesium abundances in mildly metal-poor stars from different indicators}
\author[Abia \& Mashonkina]
       {C. Abia$^1$, L. Mashonkina$^2$ \\
        $^1$ Departamento de F\'\i sica Te\'orica y del Cosmos. Universidad de Granada, 18071 Granada, Spain \\
        $^2$ Department of Astronomy, Kazan State University, Kremlevskaya 18, 420008 Kazan 8, Russia\\}
\date{                                                                                                }

\pagerange{\pageref{firstpage}--\pageref{lastpage}}
\pubyear{}

\begin{document}

\maketitle

\label{firstpage}

\begin{abstract}

We present magnesium abundances derived from high resolution spectra using
several Mg I and two high excitation Mg II lines for 19 metal-poor
stars with [Fe/H] values between -1.1 and +0.2. The main goal is to search
for systematic differences in the derived abundances between the two ionisation
state lines. Our analysis shows that the one-dimensional LTE and N-LTE study
finds a very good agreement between these features. The [Mg/Fe] vs. [Fe/H] relationship
derived, despite the small sample of stars, is also in agreement with the classical
figure of increasing [Mg/Fe] with decreasing metallicity. We find a significant scatter however,
in the [Mg/Fe] ratio at [Fe/H]$\sim -0.6$ which is currently explained as
a consequence of the overlap at this metallicity of thick and thin disk stars, which
were probably formed from material with different nucleosynthesis histories. We speculate
on the possible consequences of the agreement found between Mg I and Mg II lines 
on the very well known oxygen problem in metal-poor stars. We also study the [O/Mg] ratio
in the sample stars using oxygen abundances from the literature and find that the current 
observations and nucleosynthetic predictions from type II supernovae disagree. We briefly discuss
some alternatives to solve this discrepancy.    
\end{abstract}


\section{Introduction}
It  is  currently  assumed   that  the  chemical  composition  of  the
atmospheres of unevolved (dwarf)  stars reflects that of the material
from which the stars are formed. Therefore, abundance studies in these stars
covering a wide  range of metallicities is probably the best tool to
study the different scenarios proposed for the formation of the Galaxy,
its star formation rate history and chemical evolution.

During the  last decade there  has been a considerable  improvement in
stellar abundance  analyses.  This fact,  together  with the  more
accurate  derivation  of   the  basic  stellar  parameters  (effective
temperature, gravity etc),  mainly after the general use  of the {\it
Hipparcos  Catalogue} data  (ESA 1997),  has considerably reduced the
uncertainty in  the derivation of  stellar abundances;
formal errors  of a few hundredths  of dex in the  abundance ratios are
now being obtained!  (see  e.g. Reddy  et  al.  2003).  On  the  other hand,  the
classical LTE  abundance analysis using 1D model  atmospheres is being
progressively replaced by full N-LTE abundance analysis, in some cases
using also  3D hydrodynamical  model atmospheres (Nordlund  \& Dravins
1990; Asplund et  al. 1999; Asplund \& Garc\'\i  a-P\'erez 2001). With
such strong  tools available,  recent analyses of the  abundance ratios
([X/Fe]) as a function of the stellar metallicity ([Fe/H]\footnote{The
usual bracket notation is used  throughout the paper for the abundance
of  an  element  X  with  respect to  the  hydrogen  [X/H]$=  \rm{log}
{{(N_X/N_H)}_\star\over{(N_X/N_H)}_\odot}$, where N represents number
density.})  are  showing  a new, detailed structure in the [X/Fe] vs.
[Fe/H] trends (on occasions, unexpected) that were hidden in previous
analyses, in part due to lower quality data and larger abundance errors
(e.g. Gratton  et al.  1996; Nissen \&  Schuster 1997;  Fuhrmann 1998;
Idiart  \&  Th\'evenin 2000;  Prochaska et al. 2000; Mashonkina  \&  Gehren  2001; 
Gehren  et al.  2004; Gratton et al. 2003a,b). These new studies 
clarify theories of the Galaxy's formation (see e.g. Weimberg et al. 2002; Samland \& Gerhard 2003), 
in particular its star formation history, the timescales for the formation of the halo, thick and 
thin disks and, in some cases, could lead to a revision of the nucleosynthesis
theory in stars, mainly in type II supernova explosions.  
For instance, concerning    the   well   known    overabundances  of the   
named $\alpha$-elements (O, Mg, Si, Ca etc) with respect to Fe in metal-poor
stars,  there   are  claims  indicating  that  the  level   of  this
overabundance  at a given  metallicity is  different depending  on the
stellar  population  studied:  halo,  thick-disk  or  thin-disk  stars, but also for a 
particular $\alpha$-element. This poses a difficult challenge to our current understanding
of nucleosynthesis in stars. Furthermore, abundance  studies  in  extreme  
metal-poor  stars ([Fe/H]$<-2.5$) have revealed   very   peculiar  abundance   
ratios. For instance, there are a significant number of stars with [C,N,Mg,Si/Fe]$>1$), with
a considerable  degree of dispersion (Norris, Ryan \&  Beers 2001;  Depagne et  al. 2000;  Truran et
al.  2002; Aoki  et  al. 2002,  among  many others). This might  well provide 
information about  the level of  mixing of the interstellar  medium at
early  epochs as  well as  on nucleosynthesis  processes in  zero
metallicity stars.
 
A common characteristic of many stellar abundance analyses is that they are usually
based on the use of a small number of atomic lines for a specific element, in many cases
also in the same ionisation state. Of course, for some elements there are very few
spectral lines available in dwarf stars, in some cases just one
(e.g. the Li I at 6708 {\AA}) but in others, the use of several abundance indicators
does not always lead to agreement between them. In fact, in  this era of precision in
abundance analyses, the longstanding problem of the [O/Fe] vs. [Fe/H] relationship in
metal-poor stars still remains to  be solved. 
The {\it oxygen  problem} is that  several potential  indicators of  the oxygen
abundance  (ultraviolet OH lines,  the forbidden  [OI] 6300  {\AA} and
6363  {\AA} lines, the  high excitation lines of 7770 {\AA} O I  triplet, 
and the near infrared vibration-rotation lines from  the  
OH  molecule's  ground  state), may  give different results (see the recent 
discussions by Nissen et al. 2002, and Fulbright \& Johnson 2003, for details). 
This fact provokes some doubts regarding our current understanding of line formation in the
atmosphere of metal-poor stars and, in consequence, regarding the implications to be drawn from the abundance analyses for the important nucleosynthetic and
galactic issues mentioned above.

In this paper, we address the difficulty using an example:
magnesium. It is widely  accepted that the [Mg/Fe] ratio in metal-poor  stars
increases  for decreasing  metallicity, reaching  a  plateau-like value
$\sim  0.3-0.4$  dex  at  [Fe/H]$\leq -1$. Magnesium is  an
$\alpha$-element which is produced in  the interior of  massive stars
during the hydrostatic burning phases of the stellar evolution.   
Nevertheless, recent studies have revealed that
the [Mg/Fe] ratio shows greater scattering at a given metallicity than other $\alpha$-elements
like, Si, Ca or Ti (we exclude here O); moreover, there is a sudden increase in the [Mg/Fe] ratio at
[Fe/H]$\sim -0.6$ and/or a clear distinction of the [Mg/Fe] ratio depending on
which population the star belongs to (see e.g. Prochaska et al. 2000; Qiu et al. 2003 
and references therein). 
 
Similarly to oxygen, there are several atomic and molecular Mg lines available
to derive the abundance of  this element in dwarfs, including
two  Mg II  lines  at 7787  and  7796 {\AA} not commonly used in the literature 
and which  have a similar
excitation energy ($\chi=9.99$ eV) to those of the O I triplet ($\chi=9.14$ eV).
From the  point of view  of classical 1D  analysis, this  implies that
these  Mg II lines  should form  at essentially  (however, see below)  the same
depth in the atmosphere as the oxygen triplet lines.  Furthermore, since
neutral oxygen and  single ionised magnesium  are  both  majority
species in metal-poor dwarf stars, they are expected to be affected in a
similar  way by stellar  inhomogeneities (granulation, Asplund  2003). Furthermore,  given the
high  excitation  energy of  these  lines,  they  are expected  to  be
as sensitive to the temperature structure of the atmosphere as the oxygen
triplet. Departures  from LTE in these  Mg II lines  are also expected,
for  the  same  reason. In brief, we wonder whether analysis of these
Mg II would give abundances in agreement with those obtained with the 
more commonly used Mg I lines or, similarly to oxygen, there is
also a {\it  magnesium  problem}.

The  paper is  organised as  follows: Section  2 describes  the
observations  and the sample  of stars.  Section  3 presents the
abundance analysis  of the selected Mg  I and Mg  II lines considering
also  the corrections for N-LTE effects.  In Section  4 the  results are
compared  with   previous  studies  and  finally   our  conclusions  are
summarised in Section 5.

\section{Observations and data reduction}

The  magnesium and  iron  abundances were  derived from  spectra
obtained with the SOFIN echelle spectrograph (Touminen et al. 1998) at
the  2.5m NOT  telescope in  the  Roque de  los Muchachos  Observatory
on 9-11 January 2001.  Using the first camera and the appropriate
slit width (38 $\mu$m, corresponding to two pixels on the CCD), the resolving  power achieved
was $\sim  $170000\footnote{Due to not perfect seeing conditions and in order to gain light, for some stars observed with SOFIN 
we used a slightly larger slit. In this case the resolving power achieved was lower that the quoted value.}.  
The  spectral  range 4850-10100  {\AA}  was  observed
simultaneously although with large gaps between the different spectral
orders. Unfortunately, in the stars observed with this instrument, this limited the number 
of Mg I and iron lines available for the analysis. Therefore, comparison between Mg I
and  Mg  II  absolute  abundances  in the  stars  observed  with  SOFIN
may lack some significant  statistics: apart from the two Mg
II lines, these spectra include only the Mg I line at 5711 {\AA}. This
was  not the  case,  however, for  the  stars observed  with the  2.2m
telescope at Calar Alto Observatory  on 18-20 July 2000 and 
9-10 August 2003   using  the   echelle  spectrograph  FOCES   (Pfeiffer  et
al.  1998). In  this case  the spectra  covered the  full  range between
4400-9000 {\AA}, at the cost of the lower resolving power ($\sim
40000$) achieved. These  spectra, hence, contain numerous Mg  I and Fe
II lines although, unfortunately, the Mg II 7877 {\AA} line was placed
near the border of  the  detector, where efficiency is not
good. This meant that for some stars this line could not be used.

\begin{figure*}
\includegraphics[width=190mm]{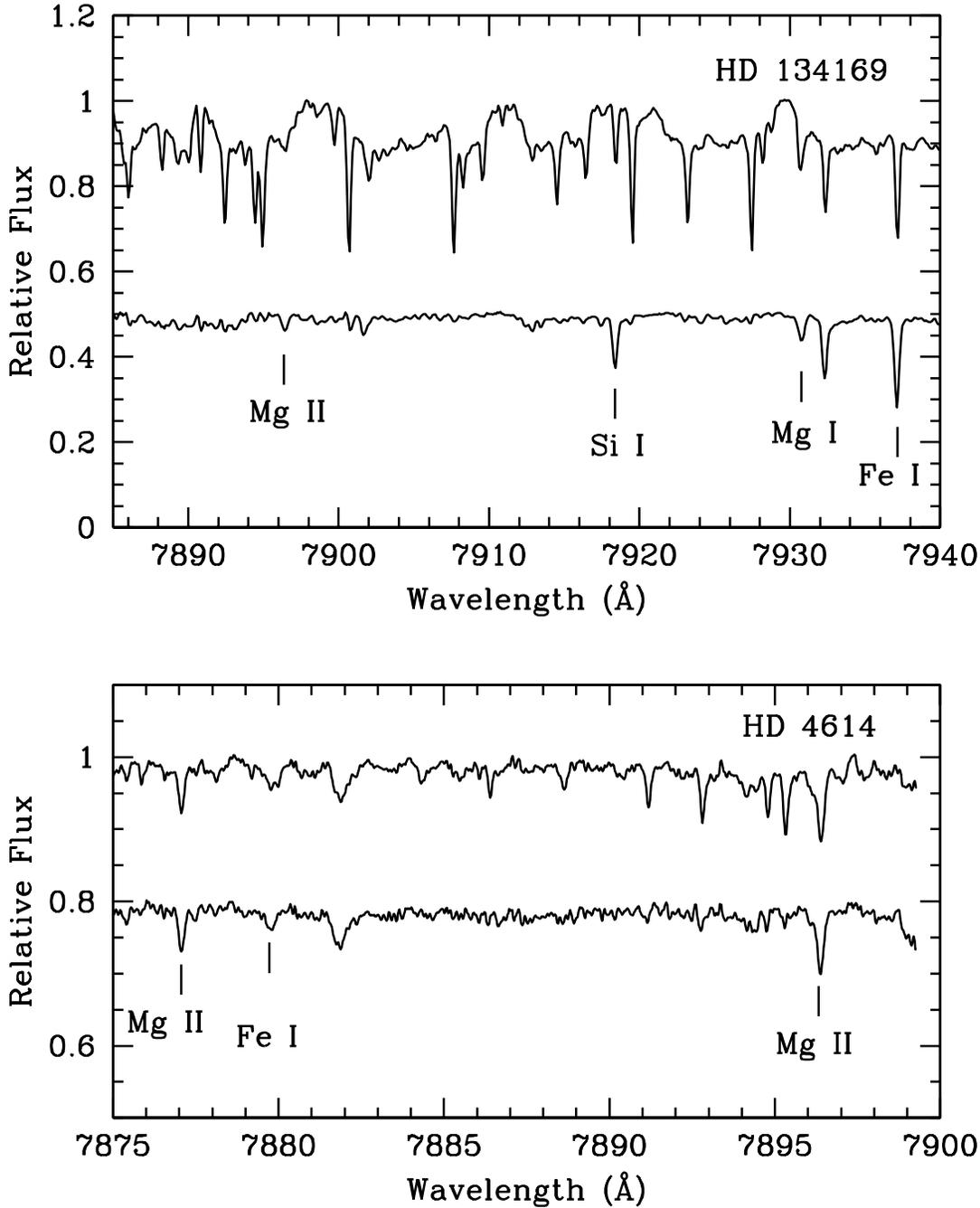}
\caption{ Representative  spectra in  the region  of the  Mg  II lines
before  removing the telluric  lines (upper spectrum)  and after (lower spectrum), respectively.
The lower spectrum  in each panel has been shifted for clarity. Note the different spectral range and
ordinate scale in the plots. The spectra shown also have different spectral resolution. The spectrum of HD 134169 was obtained
with FOCES, while HD 4614 with SOFIN. In HD 134169 the Mg II 7877.059 {\AA} line was not 
detected since due to the radial velocity of the star, the line was placed near the border of the detector. The Mg I 7930.819 {\AA}, 
however, is clearly seen in this star. This line served as a test to check the telluric line removal procedure (see text). }
\end{figure*}

The  spectra  were  reduced  using  IRAF  following standard
techniques for the extraction  and calibration of echelle spectra. The
spectra  obtained  with SOFIN typically have $S/N\sim  100$ in  the
spectral regions of the Mg I and Mg II lines. In  the FOCES spectra
the $S/N$ depends on the spectral region but in general they are good, ranging
from $\sim  50$ in the bluest part  to more than 200  at $\sim 8000$
{\AA}.  The main difficulty in the extraction of the spectra in the Mg
II  region is the  blend with  telluric H$_2$O  lines, which  are quite
numerous near  the 7877 {\AA} Mg  II line. In  order to remove  these lines,
several  rapidly  rotating B-type  stars  were  observed  as close  as
possible in the sky with respect to  the target star. We used the IRAF task
{\it telluric} to  optimise the fit between the  telluric lines of the
program  star and  the  hot star  by  allowing a  relative scaling  in
airmass and wavelength  shift. The procedure worked well in most of the
stars, but in some cases, the hot-type star could not  be observed very
close in time and in the sky  to the program star.  Small {\it spike}
features  appeared in these  cases in  the final  spectrum, probably
indicating a non-perfect scaling of the airmass. The procedure of telluric
absorptions removal certainly adds an extra uncertainty in the  derivation of magnesium
abundance from the Mg II lines. We realized that when the S/N ratio of both the target and the hot star spectra were not very high, the
clean spectrum sometimes had an ill-defined continuum, which translates into a larger error in the
equivalent width measurement of the lines. In fact, in some of the stars we derived only upper limits (see below). 
Figure 1 shows two representative spectra  in the region of  the Mg II  lines before and
after the removal of the telluric lines.

\begin{table*}
\centering
\begin{minipage}{160mm}
\caption{Str\"omgren photometry, colour excess, $(V-K)$ index and absolute 
visual magnitude derived from Hipparcos parallaxes for the sample stars. Masses
are derived from interpolation in theoretical isochrones log L/L$_{\sun}$ vs. log T$_{\rm eff}$ (see text).
The spectrograph used in the observations is indicated.}
\begin{tabular}{@{}lccccccccccc@{}}
\hline
Star\footnote{In the stars marked with an asterisk, the removal of the telluric lines in the Mg II region might be imperfect.}&   
Instrument & S/N\footnote{Average signal to noise ratio achieved in the spectral region of the Mg II lines.}&$V_o$  &$(b-y)_o$ & $m_o$  & $c_o$ & $\beta$ & $E(b-y)$ & $(V-K)_o$ & $M_{V}$ & $M/M_\odot$ \\
\hline 
HD  400 &FOCES& 80 &6.08   & 0.350 & 0.141 & 0.389 & 2.615 & $-$0.019 &...    & 3.67 & 1.20\\
HD 4614 &SOFIN & 110 &3.45   & 0.402 & 0.185 & 0.275 & 2.588 & $-$0.030 & 1.480 & 4.57 & 1.05\\
HD 19994&SOFIN &120 &5.06   & 0.361 & 0.185 & 0.422 & 2.631 & $-$0.007 & ...   & 3.31 & 0.95 \\
HD 51530&SOFIN &80&6.20   & 0.359 & 0.137 & 0.401 & 2.602 & $-$0.014 & 1.395 & 2.82 & 1.20 \\
HD 59984&SOFIN &100&5.90   & 0.360 & 0.123 & 0.336 & 2.594 & $-$0.004 & ...   & 3.51 & 1.00 \\
HD 106516&FOCES &230&6.10   & 0.331 & 0.114 & 0.334 & 2.618 & $-$0.012 & 1.203 & 4.33 & 0.90 \\
HD 116316&FOCES &80&7.67   & 0.316 & 0.120 & 0.381 & 2.630 & $-$0.012 & 1.211 & 4.06 & 1.00 \\
HD 134169&FOCES &275&7.68   & 0.373 & 0.119 & 0.312 & 2.582 & $-$0.003 & 1.426 & 3.80 & 1.10 \\
HD 150177$^\star$& FOCES&330&6.32  & 0.331 & 0.119 & 0.395 & 2.613 & $+$0.002 & ...   & 3.13 & 1.20\\
HD 165908&FOCES&660&5.04  & 0.353 & 0.135 & 0.322 & 2.611 & $-$0.002 & ...   & 4.06 & 0.90 \\
HD 170153&FOCES&214&3.58  & 0.336 & 0.146 & 0.317 & 2.611 & $-$0.025 & ...   & 4.07 & 1.05 \\
HD 192718&FOCES &158&8.39  & 0.396 & 0.144 & 0.303 & 2.573 & $-$0.015 & ...   & 4.57 & 0.93\\
HD 201891&FOCES&295& 7.37  & 0.361 & 0.094 & 0.261 & 2.586 & $+$0.001 & 1.397 & 4.62 & 0.90\\
HD 207978&FOCES &286&5.55  & 0.301 & 0.122 & 0.425 & 2.640 & $-$0.002 & 1.210 & 3.33 & 1.05\\
HD 208906$^\star$&FOCES &350&6.95  & 0.355 & 0.122 & 0.292 & 2.605 & $-$0.012 & 1.402 & 4.61 & 0.97\\
HD 210595&FOCES &192&8.60  & 0.305 & 0.101 & 0.494 & ...   &  ...   & 1.222 & 2.62 & 1.30\\
HD 210752&FOCES &195&7.40  & 0.377 & 0.132 & 0.290 & 2.585 & $-$0.016 &  ...  & 4.52 & 1.00\\
HD 215257$^\star$&FOCES &238&7.46  & 0.358 & 0.116 & 0.310 & 2.594 & $-$0.001 & 1.530 & 4.33 & 0.90\\
BD +18$^{\circ}$ 3423&FOCES&122& 9.79 & 0.357& 0.090 & 0.319 & 2.583 & $-$0.007 & 1.440 & 4.54 & 0.95\\
\hline
\end{tabular}
\end{minipage}
\end{table*}

\section{Stellar parameters and spectroscopic data}
The  near infrared Mg  II lines  are weak,  even in  solar metallicity
stars. This limited  the  metallicity range of the program  stars studied. In fact,
theoretical simulations show that for stars with [Fe/H]$<-2$, the
expected equivalent  width in the most favourable  cases (T$_{\rm eff}>
6000$ K and log g $< 4.0$) of the stronger of the two Mg II lines (7896
{\AA}), would  not exceed  $\sim 5-6$ m{\AA}, even assuming  a typical
enhancement [Mg/Fe]$\sim +0.4$ dex for a star of this metallicity. The
difficulty in detecting such a weak spectral feature is obvious, and is more so 
when it might be affected by telluric absorptions (e.g. it becomes highly
difficult  to  place  the  continuum). We  decided, hence, to
concentrate on   the    metallicity   range
$-1.0\la$[Fe/H]$\la 0.0$, although this, unfortunately, limits our conclusions.

An important issue in abundance analysis is the derivation of the
effective temperature. Ideally, one should use an indicator that is insensitive to
errors in reddening and to other stellar parameters such as gravity. 
Theoretically, the profiles of Balmer lines satisfy these conditions, but
we could not use them because they were located at the border of the CCD
in the echelle spectra, which meant that in many cases their wings were truncated.
Therefore, the effective temperature was derived from the $b-y$ and $V-K$ colour
indexes  using the  infrared  flux method  calibrations  of Alonso  et
al.  (1996). The  Str\"omgren photometry  was taken  from  Schuster \&
Nissen  (1988)  while  K  photometry  was that of Carney  (1983),  Alonso  et
al.  (1994)   and  The  Two   Micron  All  Sky  Survey   (Finlator  et
al. 2000).  According to the {\it Hipparcos}  parallaxes, the majority
of  our stars  are within  100 pc  of the  Sun, where interstellar
reddening  is negligible. In  any case,  the $(b-y)_o$  calibration of
Schuster \&  Nissen (1989) was  used to derive  interstellar reddening
excess. Reddening was only
considered for stars  with  $E(b-y)\ga  0.02$, and  the V and K magnitudes, and the $m_1$  and $c_1$
indexes were corrected according  to the relations with $E(b-y)$ given
in Nissen  et al. (2002). Photometric values are shown in  Table 1. The
T$_{\rm  eff}$  calibrations by  Alonso  et  al  (1996) present little
dependency on metallicity, and so it was necessary to estimate a metallicity  value in
advance in order to  derive  T$_{\rm eff}$.   As a  first
estimate, we took the average metallicity  value given in the compilation by
Taylor (2003). In general, this  metallicity agreed well with the
spectroscopic  value  derived from  Fe  II  lines  (see below).  When
photometry was available, we adopted the mean value of T$_{\rm eff}(b-y)$
and T$_{\rm eff}(V-K)$. Both  T$_{\rm eff}$ estimates were in agreement
within $\pm 50$ K. For the stars with no K photometry, we took the
T$_{\rm eff}(b-y)$  value. Uncertainties in  the photometry, reddening
and the calibration of the absolute flux in the infrared mainly
determine the error in T$_{\rm eff}$. Alonso et al. (1996) estimated
an  uncertainty of  $\pm  90$  K, taking  into  account both
systematic and random errors in the calibration. Thus, we adopt a conservative
error of $\pm 100$ K in T$_{\rm{eff}}$.

Gravities were derived from the classical relation

\[
\rmn{log}~g/g_{\sun}  =   \rmn{log}~M/M_{\sun}  +  4~\rmn{log}~T_{\rmn{eff}}/T_{\rmn{eff}_{\sun}} \\
+ 0.4~(M_{bol}-M_{bol_{\sun}})
\]

taking a value of $M_{bol_{\sun}}=4.75$ for the Sun.
Absolute  visual magnitudes  $M_V$  were determined  on  the basis  of
Hipparcos  parallaxes. All  parallaxes in the sample stars have an
accuracy of $\sigma(\pi)/\pi < 0.3$, and the overwhelming majority $<0.02$. To
compute $M_{bol}$, the bolometric corrections by Alonso et al. (1995)
were used. Finally, the stellar mass was derived by interpolating in
the $\rm{log}~L/L_{\sun}$ vs. $\rm{log}~T_{\rm eff}$ relation between
the  $\alpha$-element enhanced  evolutionary tracks by VandenBerg et
al. (2000). Depending on the metallicity of the star, we used the tracks
computed with different values of the helium fraction Y (from $\sim 0.24$ to 0.28).
The  precision in the theoretical derivation of the mass was estimated
in $\pm 0.05$ M$_{\sun}$. The error of the derived value of log $g$ is
mainly determined by the error in $M_V$ ($M_{bol}$) which is dominated
by the uncertainty of the parallax. The estimated final error for log
$g$ ranges from 0.03 to 0.15 dex.

The microturbulence parameter was derived using the
relation from Reddy et al. (2003) obtained by studying 33 well defined Fe I
lines in 87 dwarf stars with effective temperatures in the range 5650 to 6300
K, log $g$ from 3.6  to 4.5 and metallicity $-0.8\la$[Fe/H]$\la +0.1$. 
The metallicity of three stars in our sample is a little lower than this range, but  
we believe the expression can be safely used for
the  entire sample since the use of relatively weak lines  makes the
abundance analysis almost insensitive to errors in the microturbulence
parameter. The relation is

\[
\xi_t  =  1.28  +  3.3  \times  10^{-4} (T_{\rm  eff}  -  6000)  -0.64
~(\rm{log}~g - 4.5)
\]

The {\it rms} error in the least-squares fit derived by these authors is
$\sigma\sim 0.15$ kms$^{-1}$. Similar  linear  regressions  have  been  used  by  others
(Edvardsson et al.  1993; Chen et al. 2000),  giving similar results
within  $\pm 0.25$ kms$^{-1}$, and showing little or no dependence on
metallicity as long as weak lines are used (see also Fuhrmann 1998).
As a check, we compared our microturbulence values with those spectroscopically derived
in works with stars in common. With respect to the work by Gratton et al. (1996) we found
a difference of $+0.04 \pm 0.21$ kms$^{-1}$ (6 stars in common), $-0.08\pm 0.15$ kms$^{-1}$
with Edvardsson et al. (1993, 13 stars) and $+0.19\pm 0.11$ kms$^{-1}$ with Fuhrmann (2004, 7 stars).
Thus, we also adopt a conservative uncertainty in the microturbulence of $\pm 0.25$  kms$^{-1}$. In fact, variations
of this amount are unimportant as far as the abundance analysis  
is concerned, except for the relatively strong Fe II lines at 5197 and 5234 {\AA} in 
the near solar metallicity stars of our sample (see below). 

Table 1 also shows the stellar masses derived, and the final
values  of T$_{\rm  eff}$, log  $g$ and microturbulence are  given in
Table 2, together with the values  of [Fe/H] derived from  Fe II lines
(see  Sect. 3.1). As the calibration of T$_{\rm eff}$ and log $g$
depends somewhat on  [Fe/H], the determination of the atmosphere
parameters was an iterative process. Comparison between our estimates of T$_{\rm{eff}}$ and
gravity with those derived by authors using different methods shows good agreement. 
We found mean differences of $-13\pm 77$ K (Gratton et al.
1996), $-73\pm 67$ K (Edvardsson et al. 1993) and $-73\pm 57$ K (Fuhrmann 2004) for the effective
temperature in the sense of our values minus theirs, whereas for gravity, $-0.07\pm 0.15$ 
(Edvardsson et al. 1993), $-0.07\pm 0.15$ (Gratton et al.
1996) and $+0.025\pm 0.060$ (Fuhrmann 2004). Thus, for the adopted uncertainty in T$_{\rm{eff}}$ and log g, our
stellar parameters are well within 1 $\sigma$ with respect to the above studies.

\begin{table}
\centering
\caption{Stellar parameters}
\begin{tabular}{@{}lcccc@{}}
\hline
Star   &   T$_{\rm eff}$(K) & log $g$ &  $\xi$(kms$^{-1}$)   & [Fe/H] \\
\hline
HD  400  & 6147 & 4.15 & 1.55 &$-$0.17\\
HD 4614  & 5790 & 4.34 & 1.31 &$-$0.35\\
HD 19994 & 6030 & 3.90 & 1.67 & $+$0.18\\
HD 51530 & 5910 & 3.75 & 1.73 & $-$0.43\\
HD 59984 & 5871 & 3.91 & 1.61 & $-$0.77\\
HD 106516& 6105 & 4.40 & 1.38 & $-$0.85\\
HD 116316& 6195 & 4.24 & 1.51 & $-$0.67\\
HD 134169& 5806 & 4.00 & 1.54 & $-$1.00\\
HD 150177& 6084 & 3.91 & 1.68 & $-$0.73\\
HD 165908& 5924 & 4.10 & 1.51 & $-$0.67 \\
HD 170153& 6030 & 4.21 & 1.48 & $-$0.68 \\
HD 192718& 5787 & 4.28 & 1.35 & $-$0.66\\
HD 201891& 5842 & 4.30 & 1.35 & $-$1.10\\
HD 207978& 6270 & 4.00 & 1.68 & $-$0.66\\
HD 208906& 5875 & 4.31 & 1.37 & $-$0.76\\
HD 210595& 6256 & 3.80 & 1.81 & $-$0.64\\
HD 210752& 5847 & 4.31 & 1.35 & $-$0.65\\
HD 215257& 5890 & 4.21 & 1.43 & $-$0.72\\
BD +18$^{\circ}$ 3423& 5891& 4.31 & 1.36 & $-$0.93\\
\hline
\end{tabular}
\end{table}

\begin{table*}
\centering
\begin{minipage}{70mm}
\caption{Fe and Mg lines used in this study, showing equivalent widths 
as measured in the solar flux spectrum together with the abundances derived 
for each line}
\begin{tabular}{@{}lcccc@{}}
\hline
Wavelength & $\chi$  & log $gf$ & \multicolumn{2}{c}{Sun} \\
 ({\AA})   &   (eV)  &          &  W$_{\lambda}^{\sun}$(m{\AA})   & log $\epsilon$(X)\footnote{Abundances are given in the
scale log N(H)$\equiv 12$.} \\
\hline 
Fe II   &      &         &      &      \\
5100.66 & 2.81 & $-$4.19 & 19.6 & 7.50 \\
5197.58 & 3.23 & $-$2.28 & 88.4 & 7.40 \\
5234.62 & 3.22 & $-$2.20 & 90.5 & 7.53 \\
5325.55 & 3.22 & $-$3.27 & 43.0 & 7.50 \\
5414.08 & 3.22 & $-$3.80 & 27.0 & 7.58 \\
5425.26 & 3.20 & $-$3.42 & 44.0 & 7.60 \\
6084.10 & 3.20 & $-$3.86 & 21.3 & 7.55 \\
6149.25 & 3.89 & $-$2.77 & 38.0 & 7.54 \\
6247.56 & 3.89 & $-$2.38 & 56.2 & 7.57 \\
6416.93 & 3.89 & $-$2.79 & 42.2 & 7.67 \\
6432.68 & 2.89 & $-$3.76 & 43.0 & 7.68 \\
6456.39 & 3.90 & $-$2.13 & 66.5 & 7.55 \\
\hline
Mg I    &      &         &      &      \\
4571.096& 0.00 & $-$5.610& 110  & 7.50 \\
4730.029& 4.34 & $-$2.150& 77   & 7.49 \\
5528.409& 4.34 & $-$0.498& 300  & 7.60 \\
5711.091& 4.34 & $-$1.810& 106  & 7.50 \\
6318.750& 5.10 & $-$1.970& 43   & 7.55 \\
7930.810& 5.94 & $-$1.300& 59   & 7.65 \\
\hline
Mg II   &      &         &      &      \\
7877.054& 9.99 & $+$0.390& 15   & 7.62 \\
7896.390& 9.99 & $+$0.650& 21   & 7.60 \\
 \\
\hline
\end{tabular}
\end{minipage}
\end{table*}

\subsection{Stellar metallicity}

The  derivation of the  stellar metallicity  measured commonly  by the
[Fe/H] ratio  is a critical issue in studying the behaviour of any
abundance  ratio  [X/Fe]  vs.  [Fe/H].  Here, however, we  are  mainly
interested in the  search for systematic differences between  Mg I and
Mg II  abundances and thus any error in  [Fe/H] would have no greater consequences
than a similar systematic effect on the [Mg/Fe] ratio derived from
Mg I and Mg II lines.

Th\'evenin \& Idiart (1999) showed that departures from N-LTE of Fe I
lines may lead to systematic errors in the derivation of the
metallicity in metal-poor stars. N-LTE  metallicity corrections can
reach up to 0.35 dex, and are thus, not  negligible. However, Korn et al. (2003)
recently showed that this result depends crucially on the choice of the neutral hydrogen collision
scaling factor ($S_H$) (Th\'evenin \& Idiart did not consider this). For a value of $S_H=3$ 
and using effective temperatures derived from fitting the Balmer profiles, they found excellent consistency 
between the metallicity derived from both Fe I and Fe II lines in a small sample of metal-poor
stars. Nevertheless, the role of hydrogen collisions in keeping LTE in metal-poor atmospheres
is still a matter of debate (cf. Barklem et al. 2003) (in fact Korn et al. 2003, did not
find any physical reason to justify $S_H=3$). Thus, because iron  abundances based on  Fe II  
lines do not suffer important N-LTE effects, we based the derivation of metallicity
on the analysis of a number of relatively weak and unblended Fe II
lines. The list of lines is given in Table 3, and is almost the same
as that used by Nissen et al. (2002). The $gf$-values for these
lines are from Bi\'emont et  al. (1991) (see  Nissen et al. 2002 for
more details). To minimise systematic errors due to wrong $gf$ values,
we performed a relative analysis with respect to the Sun, line-by-line; thus,
the iron abundance in the Sun was considered as a free parameter. We
adopted the Uns\"old (1955) approximation to the van der Waals damping
constant with an enhancement of log $\gamma_6$ by a factor of 2.5. Except for the
Fe II lines at 5197 and 5234 {\AA} in the stars of the sample with largest metallicity (HD 400, HD 4614 and
HD 19994), iron  abundance is  practically independent of the damping parameter as well as  
of the  microturbulence parameter. In fact, no  
correlation was  found  between the  Fe abundance and the equivalent  widths of the lines in any star.  
By using Kurucz's model atmosphere for the Sun with parameters 5780 K,
log $g=4.44$ and $\xi_t=1.15$ kms$^{-1}$, and the equivalent widths of
the Fe  II lines  as measured  in the solar  flux spectrum  (Kurucz et
al.   1984),   the   average    solar   abundance   derived  was   log
$\epsilon$(Fe)$=7.54$, very  close to the  commonly adopted meteoritic
value of 7.50 (Grevesse \&  Sauval 1998)\footnote{Recent abundance studies in the
solar photosphere which take into account N-LTE and the effect of granulation in the
formation of Fe I and Fe II lines found a lower value, log
$\epsilon$(Fe)$=7.45\pm 0.05$ (Apslund et al. 2000; Bellot-Rubio \& Borrero 2002). 
Nevertheless, in this paper we adopt the solar abundances
cited by Grevesse \& Sauval except where explicitly mentioned.}. The relatively large 
iron abundance scatter found in the Sun ($\pm 0.08$  dex, see Table 3) is probably due to errors in the $gf$-values.
However, this problem is avoided by  working differentially  line-by-line with respect to the
Sun in the sample stars. The iron line scatter found in the program stars ranges from  
$\pm  0.07$ to $\pm  0.12$ dex. On the other hand, the error induced by the
uncertainty in T$_{\rm eff}$ ($\pm 100$ K) is typically less than $\pm 0.07$ dex, while an
error in log $g$ ($\pm 0.15$) introduces an error of about $\pm 0.06$ dex into [Fe/H]. By adding these errors 
(quadratically) to the error due to line-to-line scatter from the equivalent width measurements, we derive a 
typical error of $\pm 0.10$ dex for [Fe/H]. The error may be a little larger ($\pm 0.13$ dex) when
the 5197 and 5234 {\AA} Fe II lines are considered individually because of their sensitivity to
microturbulence.  

The  literature contains a large  number of  studies with  stars in
common, with which our  estimate of metallicity can be compared. Note that our metallicities agree to better than $\pm 0.10$ dex
with the recommended value given in Taylor's (2003) metallicity
compilation. A better criterion is to compare our results with those obtained in similar studies. Comparison with the  
N-LTE [Fe/H] values given in Th\'evenin \& Idiart (1999) shows a mean systematic difference $-0.15$  dex (in
the  sense of our results minus theirs)  which  can be  explained by  the
differences in  the  stellar  parameters  adopted (effective
temperature and gravity). With respect to the studies mentioned above, in comparing temperatures and gravities, the comparison of stars in 
common shows a mean difference in [Fe/H]
of $-0.023\pm 0.14$, $-0.04\pm 0.11$ and $-0.05\pm 0.07$ with Gratton et al. (1996), Edvardsson
et al. (1993) and Fuhrmann (2004), respectively.  Note again that  systematic errors in [Fe/H] 
would affect the [Mg/Fe]  ratios derived  from Mg  I and  Mg II
lines in the same sense. The final  metallicities derived are shown in Table 2.

\subsection{The magnesium lines}


Table 3 also shows  the list of magnesium lines used in this study. Except for
the 7930  {\AA} line,  the remaining lines  are widely used in the
literature. They  are catalogued as {\it clean} lines by  Lambert \&
Luck (1978) although  the 4730 {\AA} line could be affected by a Cr I
line at 4729.859 {\AA} and the 6318 line by the middle line belonging to
the Mg triplet between 6318-6319 {\AA}.  Due to its proximity to the Mg II
lines, the  Mg I 7930  {\AA} served as an indirect test of the
method of telluric absorption removal in this region (see below). The
$gf$-values for the Mg I lines were taken from Wiese et al. (1969) and
Kurucz \& Peytremann (1975). They agree quite well with other accurate
lifetime measurements  (Kwong et al.  1982; Lambert \& Luck  1978, and
references therein). The van der Waals damping constants adopted are those deduced by
Zhao  et  al.  (1998)  from  the N-LTE  analysis  of  several  neutral
magnesium  lines in  the solar  atmosphere, except that for the 7930
{\AA} line, where we adopted the classical Uns\"old  constant with an
enhancement by a factor of 2.5. The $gf$-value  for this line is rather
uncertain, as it  seems  to be  formed  by several (crowded) Mg  I
absorptions. In  fact, in  the  solar  spectrum  this line  shows  an
asymmetry to the  red. We used the $gf$-value deduced by Th\'evenin
(1990) from the solar spectrum. In the VALD database (Kupka et al. 1999) there appears a Ca I line
at 7930.851 {\AA} that may contribute to the 7930 {\AA} Mg I feature. However, we found
that even in the Sun its contribution is negligible ($<0.2$ m{\AA}). 

As for iron,  we preferred to keep the solar magnesium abundance free
and to carry out a line-by-line relative analysis with respect to the Sun. This should  minimise systematic errors due to uncertain
$gf$-values.  As  can be  seen  in Table  3,  using Kurucz's model
atmosphere for the Sun, the average magnesium abundance derived from
Mg I  lines is  log $\epsilon$(Mg)$=7.530\pm 0.035$  which is in good
agreement with the photospheric magnesium  abundance ($7.58\pm 0.05$,
Grevesse \& Sauval 1998). Using  the Howelger \& M\"uller (1974) model
for the  Sun, a closer figure to the photospheric value is found, namely log
$\epsilon$(Mg)$=7.568  \pm 0.019$.  We  believe this is evidence of
reliable $gf$'s. Spectroscopic data for the Mg II  lines are from the
VALD database. In this case,  we also adopted the
standard  van  der  Waals  damping  constant, multiplied  by  a factor of
2.5. Systematic differences between the Mg I and Mg II abundances due to
differences in the damping constant should be unimportant in a relative
analysis, because of the weakness of the Mg lines in our stars.  Also,
the average solar magnesium abundance obtained from Mg II lines agrees
with the solar value obtained by Grevesse \& Sauval.

Let us now discuss the effect on the abundances of
the  errors in  the   equivalent widths and  model atmosphere
parameters. Errors affecting the derived abundances of Mg vary from line to line,
although obviously this  does not put into question the validity of the LTE approach
(however, see below). Cayrel (1988) gives a method to  evaluate the error
in the equivalent widths

\[
\Delta W_{\lambda}={1.6\sqrt{w\delta x}\over S/N}
\]

where $w$  is the FWHM of the line, $\delta  x$ is the pixel size in
{\AA},  and  $S/N$  the  signal-to-noise  ratio  per   pixel  in  the
continuum. In our spectra, these parameters are different for FOCES and
SOFIN and for a given instrument, the $S/N$ ratio also varying from the
blue  orders to  the red  ones. Moreover, the error in the equivalent widths
is larger than that derived from this formula when it is applied to the Mg II
lines because of the extra uncertainty introduced in the position of the
continuum when telluric absorptions are removed. In this case, the
uncertainty of the continuum position may reach a maximum of 1-2$\%$. 
For the Mg II lines and in  the case  of FOCES  spectra, we estimate $\Delta W_{\lambda}\sim  2 - 4$ m{\AA}, while for those
obtained with SOFIN, $\sim 0.5 - 3$ m{\AA}, depending on the S/N of the spectrum. Certainly, the error is lower
for the Mg I and Fe II lines, which are not affected by telluric 
absorptions. Table 4 shows the sensitivity of the Mg abundances  
to the uncertainties in the model atmosphere for each Mg line, as well as for an uncertainty of $\sim 2$ m{\AA}
in the equivalent width as a representative case. The table is constructed using the atmosphere parameters of a 
typical star in the  sample, namely  T$_{\rm   eff}=5900$  K,   log  $g=4.00$,
$\xi_t=1.50$ kms$^{-1}$ and [Fe/H]$=-0.6$.  It is clear that the total
error is dominated by the uncertainty in the effective temperature and
gravity.  This is particularly  evident for the Mg  II lines, which
are also significantly affected by errors in the equivalent width measurement.

\begin{table*}
\centering
\begin{minipage}{160mm}
\caption{Effects on LTE magnesium abundances of changing atmosphere parameters 
for each spectral line used. A representative case is shown.}
\begin{tabular}{@{}lcccccc@{}}
\hline
line & $\Delta$ T$_{\rm eff}=\pm 100$ K & $\Delta$ log $g=\pm 0.15$ & $\Delta\xi=\pm 0.25$ km s$^{-1}$ & $\Delta$[Fe/H]$=\pm 0.10$ & 
$\Delta$ W$_\lambda=\pm 2$ m{\AA}& Total \\ 
\hline 
Mg I 4571 & $\pm 0.12$ & $\mp 0.010$ & $\mp 0.035$  & $\pm 0.010$ &$\pm 0.03$ &$\pm 0.14$   \\
Mg I 4730 & $\pm 0.06$ & $\mp 0.015$ & $\mp 0.010$  & $\pm 0.005$ &$\pm 0.04$ &$\pm 0.08$   \\
Mg I 5528 & $\pm 0.07$ & $\mp 0.070$ & $\mp 0.025$  & $\pm 0.025$ &$\pm 0.01$ &$\pm 0.13$   \\
Mg I 5711 & $\pm 0.05$ & $\mp 0.030$ & $\mp 0.020$  & $\pm 0.040$ &$\pm 0.02$ &$\pm 0.11$   \\
Mg I 6318 & $\pm 0.05$ & $\mp 0.005$ & $\mp 0.000$  & $\pm 0.010$ &$\pm 0.08$ &$\pm 0.11$   \\
Mg I 7930 & $\pm 0.04$ & $\mp 0.005$ & $\mp 0.000$  & $\pm 0.000$ &$\pm 0.08$ &$\pm 0.12$   \\
          &            &             &              &             &           &             \\
MgII 7877 & $\mp 0.09$ & $\pm 0.05$  & $\mp 0.00$   & $\pm 0.00$  &$\pm 0.15$ &$\pm 0.18$   \\
MgII 7896 & $\mp 0.08$ & $\pm 0.04$  & $\mp 0.00$   & $\pm 0.00$  &$\pm 0.15$ &$\pm 0.17$   \\
\hline
\end{tabular}
\end{minipage}
\end{table*}

The  last  column in  Table  4 indicates the total  error estimated
assuming that the  effects of all these errors on Mg abundance are
uncorrelated.  

An  important  issue is  how to compute the average
magnesium abundance derived from the neutral lines for comparison
with Mg II abundances.  We did not find any objective reason to
discard any of the Mg I lines. Thus, we decided to compute
the average  Mg I abundance by weighting  the abundances from each
line according to the total formal error shown in the last column of
Table 4, namely

\[
\rmn{log}~\epsilon({\rmn{Mg}})= 1/v \sum_i(\epsilon_i/\sigma_i)
\]

where $v$ is the statistical variance, and $\epsilon_i$ and $\sigma_i$
are the Mg abundance and the total formal error from each line (see
Table 4).  For the Mg II abundance, we derived the mean  value of the
two lines.   

Before comparing the magnesium abundances obtained with
earlier  studies and  between the  different magnesium  indicators, we
should consider the effects of departures from LTE in both the Mg I and the Mg
II lines.

\subsubsection{N-LTE effects on Mg lines}

Departures from LTE of neutral lines of Mg have been widely studied in
the literature. Starting with the analysis by Athay \& Canfield (1969)
in the solar photosphere, all the subsequent works (Mauas et al. 1988;
Mashonkina et  al. 1996; Zhao  et al. 1998;  Zhao \& Gehren  2000 etc)
agree concerning the conclusions: N-LTE effects are quite small in the Sun for
most  Mg I  lines in  optical wavelengths.  For metal-poor  dwarf stars,
departures from  LTE increase  due to photoionisation predominating over
collisions. The reduction  of gravity reduces the  efficiency of the
collisions,  and hence a  decrease in the  stellar gravity also tends to
increase N-LTE  effects. As  a rule, the lower the metallicity and gravity  and the  higher the
effective temperature, the greater the effects of N-LTE on Mg I  lines. However, for the stellar parameters of the stars studied
here,  N-LTE corrections  (positive) to Mg I abundances do not
exceed $+0.15$  dex. Unfortunately, the
literature does not contain detailed N-LTE calculations for  all the Mg I lines studied
here.  Thus,  we could  only  correct the  Mg  I  LTE abundances for cases in which
specific calculations were available. We used the N-LTE calculations by
Zhao \& Gehren (2000) for the Mg  I lines at 4571, 4730, 5528 and 5711
{\AA}, interpolating from their Table  1 for the stellar parameters of
our stars. The Mg I N-LTE corrections by Zhao \& Gehren are based on the
assumption of a common turbulence. However, when the corrections are not
very large, as is the case in question, they do not depend on the turbulence
(at least within 0.01-0.02 dex). Our N-LTE Mg I abundances
are derived using only the N-LTE abundances from these four lines, as indicated above.

To the best of our knowledge, the literature does not contain N-LTE studies of the
Mg  II lines  in  cool  stars. However,  several  analyses of  kinetic
equilibrium of Mg II exist for hot stars (Mihalas  1972; Snijders \&
Lamers 1975;  Sigut \&  Lester 1996; Przybilla  et al. 2001).  Here, we
make use of the main results of these studies. In the atmosphere of
cool stars,  Mg II is the majority species  ($N$(Mg~II)/$N$(Mg~I) $\ge$
10$^3$). Two useful consequences follow from
this; first, bound-bound ($b-b$) and bound-free ($b-f$) transitions in
Mg I have a negligible effect on  the kinetic equilibrium of Mg II and thus,
Mg I levels can be ignored in the atom model. In any case, here we 
considered five  of the lowest terms of Mg  I only for number
conservation. Second, as shown in previous studies where ions comprised
the majority species  (e.g. Ba II, Mashonkina et al.  1999; CI, NI, and
OI, Takeda  1994), departures from  LTE for Mg  II are expected  to be
caused mainly by  radiative $b-b$ transitions. Thus, one  may disregard the strong coupling of Mg II to the Mg III continuum. For
this reason, we include all  the levels of  Mg II
up to $n = 12$ in the atom  model, leaving a gap of 0.3 eV before the Mg III continuum.
All  levels  with $l\ge  4$  were  further  collapsed into  the  $^2$G
levels. Levels for  $n = 11$ and 12 were  treated as single hydrogen
levels. The multiplet  structure of all levels was  ignored. The final
atom model  includes 35  levels of Mg  II and  the ground state of Mg
III.  We used  laboratory energy levels collected by Sigut  \& Lester
(1996) as shown in their Table 1.

The oscillator strengths for the 284 allowed transitions were compiled
from a  variety of sources.  For the $3s  - np$ series,  $n = 3  - 6$,
$f-$values were  taken from the close-coupling  calculations by Butler
et al. (1984).  For transitions between the states $3p,  4s, 4p, 5s, 5p$
and  $6s$, we  adopted the  results of  Hibbert et  al.  (1983).  For the
remaining transitions between states with  $n\le 10$ and $l \le 3$, the
data from the TOPBASE  database (Cunto et al. 1993) were  used. The
remaining  $f-$values were  computed using the  Coulomb approximation
following Oertel \& Shomo (1968). The close-coupling results of Butler
et al. (1984) were adopted for the photoionisation cross section
of  the  Mg  II  ground  state. The  remaining  photoionisation  cross
sections for $l\le 3$ were taken from Hofsaess (1979). Sigut \& Lester
(1996) compared  these data  with the  recent Opacity  Project results
available through the TOPBASE database  and found that the low excited
level cross sections agree within $10\%$ at threshold. Considering the
insignificant role played by $b-f$ transitions in producing departures
from LTE for  Mg II (see below), such uncertainty  does not affect the
final results. The hydrogenic photoionisation cross sections with $Z =
2$ were adopted for $l\ge 4$.

In cool stars the ratio  of electrons to hydrogen atoms determines the
importance of both types of inelastic collisions. Drawin's (1968)
formula, as described by Steenbock  \& Holweger (1984), is widely used
to  calculate  hydrogenic  collisions,  and  it  suggests  that  their
influence is comparable with that of electron impacts\footnote{However, a recent
study of the inelastic collision of Li + H (Barklem et al. 2003) shows that
Drawin's formula typically greatly over-estimates the collisions with
hydrogen.}. In our calculations
we  take into account both types of collisions. Collision strengths
$\Omega$ for electron  excitation among the lower states of Mg II, up
to and including $5p$,  were taken from close-coupling calculations of
Sigut  \&  Pradhan  (1995).  The  remaining  collisional  rates  were
calculated  using the  formula of  van Regemorter  (1962)  for allowed
transitions, and that by Allen (1973) with $\Omega = 1$, for forbidden
ones.  For inelastic  collisions  with neutral  hydrogen,  we use  the
formula of  Steenbock \& Holweger (1984). Since  this formula provides
only  an   order  of  magnitude  estimate,   the  cross-sections  were
multiplied  by appropriate scaling factors  $S_{\rm H}$  in  order to
produce the  best fit to the solar Mg  II level populations and line
profile of the Mg II 7896 {\AA} line\footnote{This is the stronger feature of the
two Mg II lines and, in the solar spectrum, is less blended by telluric absorptions.}.
Collisional  ionisation  by  electrons  was  treated
following Seaton's formula as described by Mihalas (1978), in which the
collisional  ionisation  rate  is  taken  to be  proportional  to  the
photoionisation cross section.

The Mg  II kinetic equilibrium was calculated using the  code NONLTE3
(Sakhibullin 1983), based on the complete linearization method
described by Auer \&  Heasley (1976). The method of calculation was
up-dated including a new opacity package. In the new version, we used
the formulae and  atomic data from ATLAS9 by  Kurucz (1998) to compute
$b-f$  and $f-f$  transitions of  H, H$^-$,  H$^+_2$ and  He  I, $b-f$
transitions of neutral atoms and the first ions of the most abundant
elements, the H  I line opacity, Rayleigh scattering and electron
scattering. Furthermore, the background opacity includes the molecular
absorption  from  11 of the most abundant molecules and line
absorptions. The line opacity due to the metals was 
explicitly included in the calculations. The metal line list
was extracted from the Kurucz (1998) compilation  and contains
about  152000 lines of neutral atoms and the two first ionisation
stages between 91.2 and 10000 nm. Wavelengths of the Mg II $b-b$
transitions are not calculated from level energies but are given in
correspondence with the metal line list to correctly account for the
background line opacity.  The radiative $b-b$ rates were treated with
a Voigt profile. The  natural width of each line was computed using the
classical damping  constant, while the van der  Waals damping was obtained using the
values indicated above. The
Stark width for the resonance transition $3s - 3p$ was taken from the
close-coupling  calculations of Barnes (1971) and for the remaining
transitions we used the formulae $\gamma_4 = 2\times 10^{-8} \, n_e \,
n_{eff}^5$, where  $n_e$ is the electron number density and $n_{eff}$
is the effective quantum number.

\begin{figure}
\includegraphics[width=85mm]{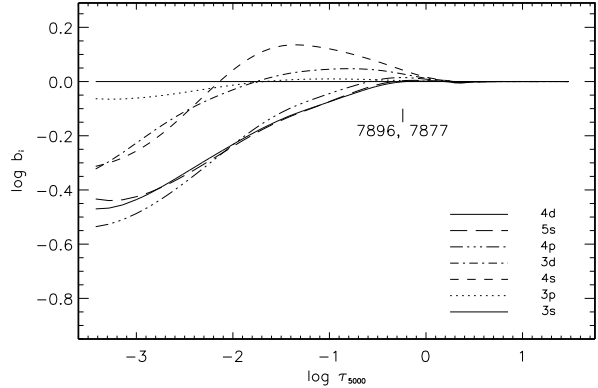}
\caption[]{Departure coefficients  $b_i$ for  some levels of  Mg~II in
the model atmospheres with T$_{\rm{eff}}=  6105$ K, log $g = 4.23$ and
[M/H]$=  -0.85$. Tick  marks indicate the locations of line centre
optical depth equal to unity for the Mg II lines at 7877 and 7896
{\AA}.}
\label{bfac}
\end{figure}

For T$_{\rm{eff}}$  and log  $g$ values close to the solar  ones, the
kinetic equilibrium of Mg II is mainly affected by radiative processes
in $b-b$ transitions, because this is the dominant ionisation state. In
Figure  \ref{bfac}, the departure coefficients, $b_i  =  n_i^{\rm
N-LTE}/n_i^{\rm LTE}$, of the few important Mg II levels as a function
of continuum optical depth $\tau_{5000}$ are shown for  a typical
atmosphere of the star sample (T$_{\rm{eff}} =  6105$ K, log  $g =
4.23$  and [M/H]$=  -0.85$). Here, $n_i^{\rm  N-LTE}$ and  $n_i^{\rm
LTE}$  are the  N-LTE and thermal (Saha-Boltzmann) number densities,
respectively. No process seems to affect the Mg II ground state population, and $3s$
maintains its  thermodynamic value. The departure coefficients of all the
excited  levels begin to deviate from 1  at depths around  log
$\tau_{5000}= 0.25$, where photon losses in the  resonance line wings
start to become important. Overpopulation of the excited levels
$3p, 4s$ and  $3d$ inside log $\tau_{5000} = -1.5$  is produced by the
pumping transitions, the resonance  transition, $3s -  3p$ ($\lambda
2800$ {\AA}), in  the first turn,  $3p - 4s$  ($\lambda 2936$ {\AA}) and $3p  - 3d$
($\lambda 2798$ {\AA}). Enhanced  excitation of  the levels  above  $3d$ is
provided by the close coupling to $4s$ and $3d$ and to each other. The
departure  coefficients of excited levels start to decrease in the
upper layers (log  $\tau_{5000}  < -1$), which  are transparent  with
respect to the radiation of the subordinate lines. From the behaviour
of the departure coefficients, we expect the lines at 7877 and
7896 {\AA} of the multiplet $4p  - 4d$ to be  amplified, compared with
the LTE case. In the line formation layers $b_{4d}  < b_{4p}$ holds,
and the source function approaches $S_{4p,4d} \approx b_{4d}/b_{4p} \
B_\nu(T_e) <  B_\nu(T_e)$. In addition, the departure coefficients of
the lower level are larger than 1, which strengthens the lines studied
here.

Since collisions with neutral  hydrogen could affect the kinetic equilibrium
of Mg II (see, however, Barklem et al. 2003), we have tried  to find their efficiency empirically from the
fit  to  the profile  of  the  Mg II  7896  {\AA}  line  in the  solar
spectrum.  For  the  Sun,  we compared different atomic models
excluding and including collisions with neutral hydrogen using various
factors, $S_{\rm  H} = 0.005, 0.01, 0.1$  and 1. Figure  3 shows
the  results  for the  Mg  II 7896  {\AA}  line.  All the  theoretical
profiles were calculated using  the oscillator strengths and Mg
abundances listed in Table  3. The theoretical profiles were convolved with a
profile that  combines rotational and  macroturbulence broadening with
velocities  $V_{rot}=1.8$ kms$^{-1}$ and  $V_{mac} =  4.2$ kms$^{-1}$,
respectively. For the sake of comparison, the LTE profile corresponding to the same
fitting parameters is also shown. Unfortunately, as noted above, this Mg II
line  is located in the  red wing of a  telluric absorption and our
calculations cannot  reproduce the continuum level correctly (see
Figure 3).  If hydrogen collisions are small in number ($S_{\rm H} =
0.005,  0.01$) we  obtain slightly broader and  deeper theoretical
profiles compared with the  LTE case. On the contrary, the inclusion of
these processes with  $S_{\rm H} = 0.1$ and 1  makes the N-LTE profile
shallower and  narrower than  the LTE one.  From Figure 3 we conclude
that $S_{\rm H}$  should be lower than 0.1. In our previous
studies, a value of  $S_{\rm H} \sim 0.01$ was derived from analysis
of the 10327 {\AA} Sr II line (Mashonkina \&  Gehren 2001). The
present analysis does not contradict the earlier ones and therefore
we accept $S_{\rm H} = 0.01$. 

\begin{figure*}
\includegraphics[width=180mm]{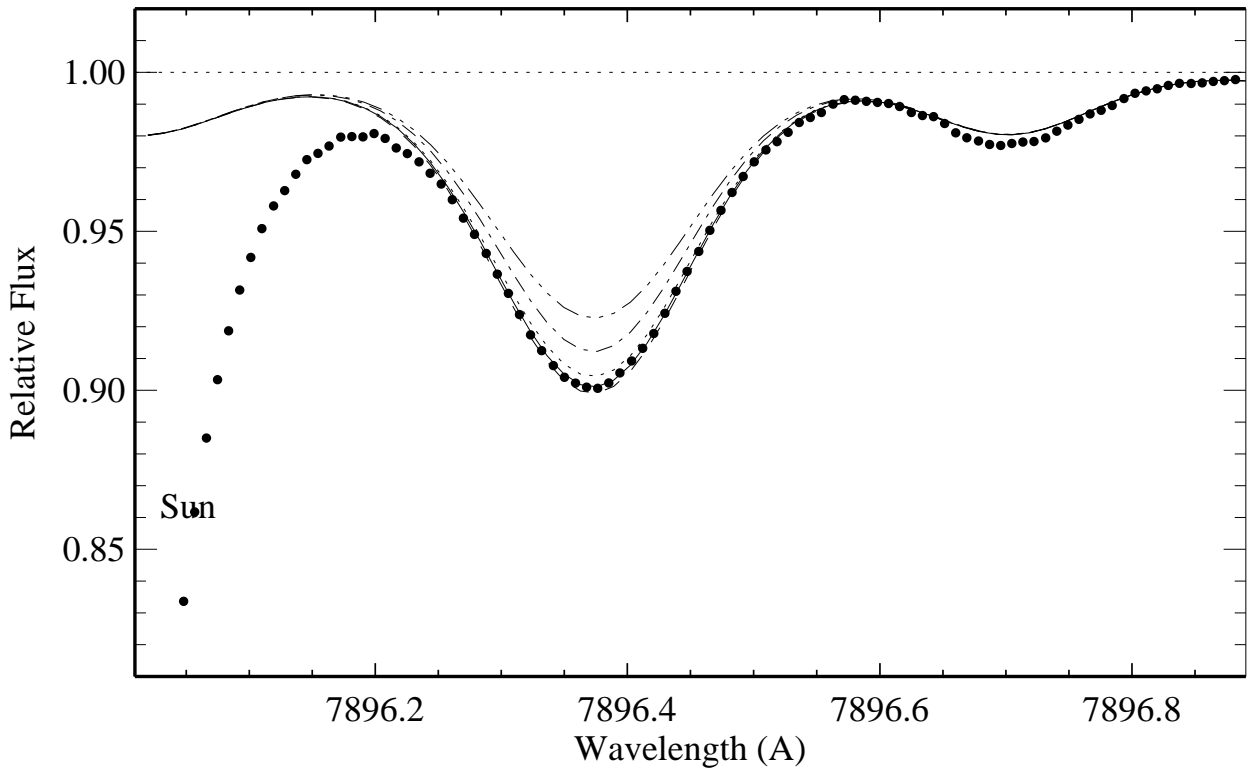}
\caption[]{Synthetic N-LTE and LTE flux profiles for the Mg II 7896 {\AA}  
line compared with the solar flux spectrum (Kurucz et al. 1984) (bold dots). 
Different lines correspond to LTE (dotted line), $S_{\rm H}= 0.005$ (dashed line), 
$S_{\rm H} = 0.01$ (solid line), $S_{\rm H} = 0.1$ (dashed-dotted line) and 
$S_{\rm H} = 1$ (dashed-three-dotted line). See text for 
discussion of the fitting parameters.}
\label{sol7896} 
\end{figure*}

For  all the  stars  in  our sample  the  N-LTE abundance  corrections
($\Delta_{\rm N-LTE}$) were calculated for both Mg  II lines as the
difference  between  N-LTE  and  LTE   abundances. As  expected from the behaviour of the departure
coefficients, the two  Mg II lines are strengthened compared with the
LTE  case  and, for the whole range of stellar
parameters investigated, $\Delta_{\rm  N-LTE}$ is negative. Since,  qualitatively
speaking, these lines form in quite deep layers, the N-LTE effects are
not very large and $\Delta_{\rm N-LTE}$  does not exceed 0.07 dex for
the 7877 {\AA} or 0.11  dex for the 7896 {\AA} line, respectively.
In   the   Sun,   $\Delta_{\rm   N-LTE}=  -0.02$   and   $-0.03$   dex,
correspondingly, with $S_{\rmn H}=0.01$. 
In order to evaluate the uncertainty in our N-LTE corrections, for the Sun, we performed test calculations to
check the sensitivity when electronic collision rates were varied. Increasing the van Regemorter
rates by a factor of 10 with  $S_H=0.0$ has a negligible effect (between $0.004-0.006$ dex).
Finally, since in the atmospheres of cool stars Mg II is a dominant specie, uncertainty in
the photonionization cross-section has almost no effect on the final corrections. Note however,
that if neutral H collisions had no effect at all on the kinetic equilibrium of Mg II (i.e. $S_H=0.0$), 
N-LTE corrections would be more negative. Thus, we consider our N-LTE corrections as lower limits.



\section {Results and discussion}

The abundances from Mg I and Mg II lines were derived by the classical
method  of  interpolating the  measured  equivalent  width  in  a
theoretical growth curve. Such growth curves 
were constructed for each line and star from a model atmosphere with the parameters in Table
2 interpolated in the grid of models by Kurucz (1998). The agreement
between the different Mg I features is good. For stars where more than
two Mg  I lines are used (15 stars), we derived a  mean dispersion of $\pm 0.09$ dex. This value is of the same order as the formal error
expected in  the abundances derived from individual Mg I lines (see Table
4). Considering this, we adopted $\pm 0.12$ dex as the typical error in the
[Mg I/H] ratios.

Barbuy et al. (1985) and Magain (1989) found discrepancies between the
magnesium abundances derived from the resonance intercombination line
(4571  {\AA}) and higher  excitation Mg  I lines  in giant  and dwarf
stars. They suggested N-LTE effects due to overionisation as the cause
of these differences,  which could reach $+0.6$ dex. 
However, Fuhrmann  et al.  (1995)  and  Carretta  et
al.  (2000) found  a good  agreement between the different Mg I
line  indicators.  The  same   is true when N-LTE
corrections are applied, and hence the differences in Mg abundances obtained  by
Barbuy et al. (1985) and Magain (1989) when using the 4571 {\AA} line
must be attributed to another  factor. Carretta et al. (2000) suggest
the choice of the  oscillator strength. We did
not find any systematic difference between  the intercombination line
and  the  other  higher  excitation  Mg I lines.  For  instance,  the  mean
difference between  the intercombination line
and the  mean value derived  from the 4730,  5528 and 5711 {\AA}  
lines is  $+0.05 \pm  0.05$. No correlation  was found either between
this abundance  difference and T$_{\rm  eff}$ and/or  log $g$. In
the same line, we found no systematic differences between the Mg I
abundances derived from the  different high excitation lines. The mean
difference between  the 7930-based abundance  and that from  the 4730,
5528, 5711 and 6318 {\AA} is $+0.02\pm 0.08$. Again, no correlation was
found against  the stellar  parameters.  This would indicate 
that systematic errors in the derivation of the stellar parameters in
our  stars are probably minimal. Surprisingly, however, we did find a systematic
difference between the 5711 {\AA} line and the other two lines with equal excitation energy,
namely the 4730 and 5528 {\AA} lines. This systematic difference is $+0.09$ dex in the sense of
the 5711 {\AA} minus the mean value of the other two. We have not found an obvious
explanation for this fact. The increase in the T$_{\rm eff}$ values would help eliminate this difference, but not by very much as these three lines
show a similar behaviour against T$_{\rm eff}$ variations (see Table 4). We searched for
possible unidentified blends at 5711 {\AA} in the data bases of Kurucz and
VALD, but no line was found that might produce such an effect. Nor are N-LTE effects
the solution, as these are similar for the three lines
and are found in the same sense (see Zhao \& Gehren 2000). Although this
systematic difference is small considering the formal error bar, it merits further study 
as many works in the literature base their Mg abundances only on the 5711 {\AA} line. 
On  the other hand, the fact
that the Mg I abundances derived from the 7930 {\AA} line agree with
those obtained from the other Mg I  lines gives us some confidence in the method
used to remove telluric lines, which may affect the Mg II lines.

Let us now compare our results with those reported from other magnesium  abundance  analyses  in  the
literature. We  have twelve  stars in common  with the ample study by
Edvardsson  et  al.  (1993) in  field  dwarf  stars.  Our Mg  I  based
abundances  are  systematically  lower by  a factor  $+  0.10  \pm  0.12$
dex.  This  can be  explained,  however,  by  the differences  in  the
effective  temperature  and  gravity  adopted.  These  authors  derive
systematically  higher temperatures  and gravities  than found by us (see above).  
From Table 4, these  differences in  the stellar
parameters easily account for the differences in Mg  abundances. In
fact, for the  four stars in common with these authors where the
differences in T$_{\rm eff}$  are less than 50  K, the  Mg abundances
agree  to better than $\pm  0.05$  dex. We also found agreement
between  error bars with  other similar  studies, namely mean  differences of
$-0.01$ dex  with Fuhrmann et al.  (1995) (4 stars),  $+0.07$ dex with
Carretta et  al. (2000) (2 stars), $+0.10$ dex  with Mashonkina et
al.  (2003) (1 star) and $-0.08$ dex with Gratton et al. (2003a) (6 stars). In all
cases, the differences can be explained as being due to differences in the
stellar parameters.

\begin{table*}
\centering
\begin{minipage}{150mm}
\caption{Average LTE and N-LTE magnesium abundances in the program stars from Mg I and Mg II lines.}
\begin{tabular}{@{}lcccccc@{}}
\hline
Star  &   [MgI/H]$_{LTE}$ &  [MgI/H]$_{N-LTE}$& [MgII/H]$_{LTE}$&  [MgII/H]$_{N-LTE}$\footnote{Note that the
abundances shown are those obtained after considering the N-LTE corrections in the Sun.}
& $\Delta$[Mg/H]$_{LTE}$ & $\Delta$[Mg/H]$_{N-LTE}$\\ 
\hline
HD 400   &   $-$0.09      &          $-$0.04    &        $-$0.09     &          $-$0.15     &                  $+$0.00 &            $-$0.11\\
HD 4614  &   $-$0.40      &          $-$0.35    &        $-$0.42     &          $-$0.43     &                  $-$0.02 &            $-$0.08\\
HD 19994 &   $+$0.17      &          $+$0.24    &        $+$0.06     &          $+$0.00     &                  $-$0.11 &            $-$0.24\\
HD 51530 &   $-$0.32      &          $-$0.25    &        $-$0.39     &          $-$0.44     &                  $-$0.07 &            $-$0.19\\
HD 59984 &   $-$0.49      &          $-$0.43    &        $-$0.41     &          $-$0.43     &                  $+$0.08 &            $+$0.00\\
HD 106516&   $-$0.52      &          $-$0.47    &        $-$0.57     &          $-$0.60     &                  $-$0.05 &            $-$0.13\\
HD 116316&   $-$0.59      &          $-$0.50    &        $-$0.59    &          $-$0.62     &                  $+$0.00 &            $-$0.12\\
HD 134169&   $-$0.56      &          $-$0.49    &        $-$0.35     &          $-$0.39     &                  $+$0.19 &            $+$0.10\\
HD 150177&   $-$0.74      &          $-$0.65    &       $<-$0.60     &         $<-$0.64     &                 $<+$0.14 &           $<+$0.01\\
HD 165908&   $-$0.63      &          $-$0.57    &        $-$0.60     &          $-$0.62     &                  $+$0.03 &            $-$0.05\\
HD 170153&   $-$0.52      &          $-$0.43    &        $-$0.50     &          $-$0.53     &                  $+$0.02 &            $-$0.10\\
HD 192718&   $-$0.44      &          $-$0.38    &       $<-$0.40     &         $<-$0.42     &                 $<+$0.04 &           $<-$0.04\\
HD 201891&   $-$0.82      &          $-$0.79    &       $<-$0.65     &         $<-$0.65     &                 $<+$0.17 &           $<+$0.14\\
HD 207978&   $-$0.59      &          $-$0.49    &        $-$0.45     &          $-$0.50     &                  $+$0.14 &            $-$0.01\\
HD 208906&   $-$0.73      &          $-$0.64    &       $<-$0.50     &         $<-$0.51     &                 $<+$0.23 &           $<+$0.13\\
HD 210595&   $-$0.52      &          $-$0.47    &        $-$0.52     &          $-$0.60     &                  $+$0.00 &            $-$0.13\\
HD 210752&   $-$0.59      &          $-$0.57    &        $-$0.47     &          $-$0.49     &                  $+$0.12 &            $+$0.08\\
HD 215257&   $-$0.66      &          $-$0.59    &       $<-$0.60     &         $<-$0.61     &                 $<+$0.06 &           $<-$0.02\\
BD +18$^\circ$ 3423&$-$0.94&         $-$0.91   &        $-$0.97     &           $-$0.98     &                  $-$0.03 &            $-$0.07\\
        &                &                     &                    &                       &                          &                   \\
Average &                &                     &                    &                       &                  $+$0.021 &            $-$0.077 \\
Dispersion&              &                     &                    &                       &                $\pm 0.081$ &         $\pm 0.091$  \\
\hline
\end{tabular}
\end{minipage}
\end{table*}

\subsection{Abundances from Mg II lines}

Table 5  shows the  final [Mg/H] ratios  derived from  the Mg I and  Mg II
features. The ratios corrected for N-LTE effects according to Section
3 are also shown. Columns 6 and 7 compare LTE and N-LTE Mg abundances
in the  sense $\Delta$[Mg/H]$=$ [MgII/H]$-$[MgI/H],  respectively.  As
can  be seen from column  6, the classical LTE abundance analysis
gives a very good agreement between the Mg I and Mg II lines, the mean
difference being only $\sim +0.02$ dex, although with a significant scatter
($\sim\pm 0.08$ dex) which is, however, compatible with the expected formal error
in the Mg I-II  abundances. When applying N-LTE corrections, agreement
is also found  between Mg I and Mg II  although, surprisingly, the mean
difference  is  now larger  ($\sim -0.08  \pm  0.09$  dex) and  changes  of
sign\footnote{Upper limits in Table 5 are not considered when deriving
differences between Mg I and Mg II-based abundances.}.  We believe
the larger difference between Mg  I and  Mg II  N-LTE  abundances is
in some way artificial. The  question is that our estimated formal error in Mg I
and Mg II abundances (see Table 4) in individual stars is, in
most cases, much  larger than the corresponding N-LTE abundance
correction  (see  Sect.  3).   Therefore,  in  this  particular  case,
considering departures from N-LTE  effects does not necessarily improve the
agreement between Mg  I and Mg II but introduces  some {\it noise}. 
[Note that ignoring the collisions with neutral hydrogen ($S_H=0.0$) in the
N-LTE calculations on Mg II would slightly enlarge the difference between
Mg I and Mg II in the N-LTE case]. 

In order  to test for systematic differences  between  Mg  I  and  Mg  II
abundances, we searched for correlations with the stellar
parameters. Figure 4 shows
that there is no apparent correlation with T$_{\rm  {eff}}$ and/or
gravity. Obviously, this should be tested further using a larger sample
of stars, spanning a wider range in metallicity.
  
Figure 5 plots the
typical [Mg/Fe] vs. [Fe/H] relationship obtained from Mg I and Mg II
lines. From the  discussion above, it is obvious that these diagrams are
very similar. Near solar ratios ([Mg/Fe]$\approx 0.0$) are obtained in
stars with [Fe/H]$\ga -0.6$, while for stars with a lower metallicity, a
small  enhancement is typically  found  ([Mg/Fe]$>0.0$).  Note  the
increase  and/or scatter  of the  [Mg/Fe] ratio  near  [Fe/H]$= -0.6$,
independently of the Mg  indicator.  This  effect was also found in
other magnesium studies (see references in Sect. 1). Fuhrmann et  al. (1995) suggested  
that this is a consequence of the onset of the
Galactic  disk  formation.   More  extended  analyses  (e.g. Edvardsson  et
al. 1993;  Prochaska et al. 2000; Reddy et  al. 2003)  found a fairly similar feature  in the
[Mg/Fe] ratio at this metallicity,  although not as abrupt as that found
by Fuhrmann et al. Indeed, differences in the [Mg/Fe] ratios are found
in metal-poor  stars with  overlapping metallicity, which  is currently
interpreted as  differences in the star formation history between the
Galactic halo, thick and thin  disk and/or due to chemical inhomogeneities
in  the  interstellar  medium  (see e.g. the critical discussion by Fuhrmann 1998). Similar  differences
between  different population stars at a  given metallicity seem also to
exist in the [O/Fe] ratio (Bensby et al. 2003) and in other $\alpha$-elements,
such as Ca and Ti (Prochaska et al. 2000). In  our small
sample, this effect might also be present. In fact, according to their
kinetic properties (T. Borkova, private communication) the stars HD
106516,  HD  134169, HD  192718,  HD 201891  and  HD  208906 in our sample might  be
considered as members of the thick  disk. The remaining stars, except BD
+18$^\circ$ 3423 (a halo star; $(U,V,W)=(-83,-264,-50),
V_{\rm{pec}}=282$ kms$^{-1}$,  $e=0.89$), probably belong to the thin
disk.  At the overlapping metallicity of the thick and thin disks
([Fe/H]$\sim  -0.6$), thick  disk stars tend to have higher [Mg/Fe]
ratios than thin disk stars (Fuhrmann 1998; Gratton et
al. 1999). Comparison of the [Mg/H] ratio (see Table  5) in the thick
disk  star HD  192718 with those of thin  disk stars of similar
metallicity  (e.g. HD  170153  and  HD  210752)  supports   with  this
statement. Thus, at  least in part, the scatter  in the [Mg/Fe] ratios
that we found in the range  $-0.7\la$[Fe/H]$\la -0.5$ may be due to
this fact (see Figure 5).

\begin{figure}
\includegraphics[width=105mm]{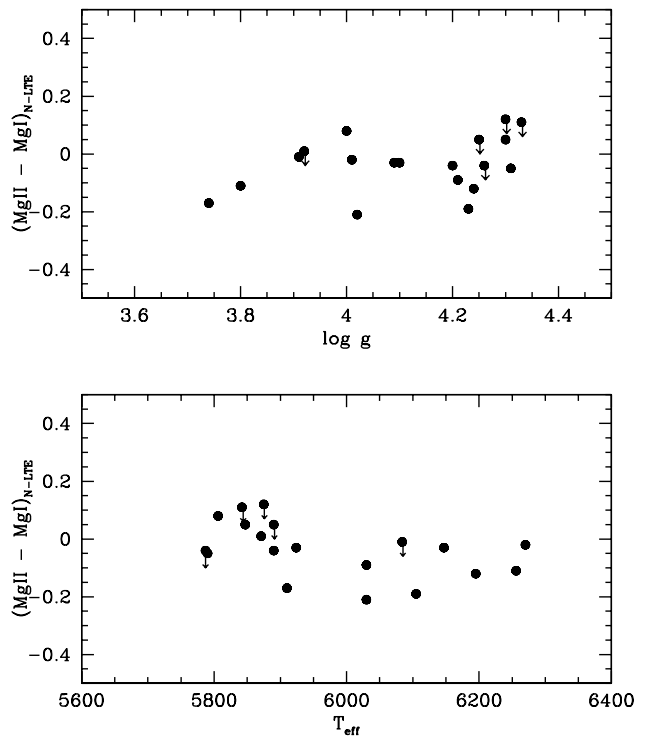}
\caption{Difference between the N-LTE magnesium abundances from Mg I and Mg II lines against 
the effective temperature and gravity of the stars. No correlation is seen.}
\end{figure}

\begin{figure}
\includegraphics[width=90mm]{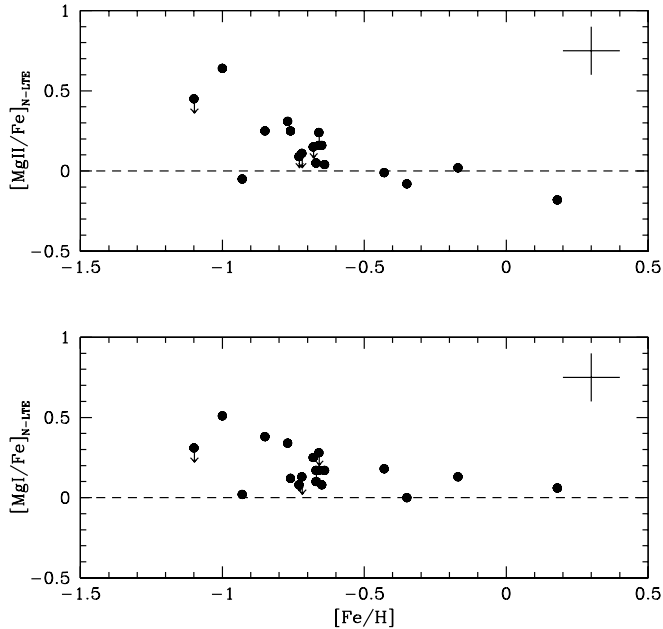}
\caption{[Mg/Fe] vs. [Fe/H] relationship in the sample stars from Mg I and Mg II N-LTE abundances.}
\end{figure}

\subsection{Magnesium and oxygen}

To finish this section we shall use the results described above
to address the issue of the oxygen problem in metal-poor stars in an indirect way.  We found a good agreement
between  the Mg  abundances  derived from  neutral  lines of different
excitation energies and those derived from Mg II lines, which have
a similar excitation energy to that of the O I triplet. When N-LTE
corrections are applied to Mg abundances, the agreement between the different Mg
indicators  remains.   In  consequence, the  [Mg/Fe]  vs. [Fe/H]
relationship obtained from the Mg  I and Mg  II lines is essentially the
same and agrees with the widely accepted behaviour of $\alpha$-elements
in metal-poor stars. This would indicate that when the classical
1D analysis is used, these Mg  II lines can be considered as {\it good
indicators} of the Mg abundance in metal-poor stars, at least in the
metallicity range studied here. Let us now consider the consequences of this
result with respect to the oxygen triplet-based abundances.

During the last decade, several approaches have been used to solve
the oxygen problem (see an update of the present status in e.g., Nissen et al. 2002 or 
Fullbright \& Johnson 2003). Fisrt, it is well known that the formation of the O
I triplet  in late-type stars is affected by departures
from LTE. The higher the temperature and the
lower the  gravity of the star, the greater is this effect. However, detailed  N-LTE
computations  (e.g. Eriksson \&  Toft 1979;  Takeda 1992,  1994, 2003;
Kiselman 2001)  have repeatedly shown that N-LTE corrections are not
enough make the oxygen abundances derived from the
triplet  agree with those obtained from  the [OI] features\footnote{By the way,  we
should comment that,  although with a less complete O  I atom
model, N-LTE effects on the  triplet were considered by Sneden
et al. (1979) and Abia \&  Rebolo (1989). These authors reached the
same conclusions as the most recent and detailed N-LTE studies.}. 

On the other hand,  it has also been argued that because of their {\it high
excitation energy} ($\chi= 9.14$ eV)  the oxygen triplet forms deep in
the  photosphere  where  {\it  inhomogeneities due to granulation} are
expected to be large (obviously, granulation cannot be considered in
the  standard  1D model  atmosphere). In Table 6, we  compare the average formation depth of the Mg II
lines with that of the two  weakest lines of the oxygen triplet (those
with a similar  strength to that of the Mg  II lines) in  three stars of our
sample that cover the range of stellar parameters studied here. As an
illustration, the last line in Table  6 shows the case of a star with
stellar  parameters T$_{\rm{eff}}=6000$  K, log  $g=4.0$  and [Fe/H]$=
-2$.   [The  oxygen problem  is mainly found in  stars with
[Fe/H]$\la -1.5$ where, unfortunately, it is very difficult to detect
the  Mg II  lines]. The  calculations  were made  assuming an  oxygen
enhancement  of [O/Fe]$\sim  +0.5$, which  is typical  for a  star with
[Fe/H]$\sim -1$. Although the concept of formation depth is quite
ambiguous, since a single line may sample very different layers of the
atmosphere (Ruiz-Cobo \& del Toro Iniesta 1994; S\'anchez Almeida et al. 1996),
in the case of weak spectral lines it might still be used for a qualitative
discussion. On this basis, Table 6 shows that the O I triplet and the Mg II
lines do not differ significantly in the depth of formation. The
oxygen lines form in slightly higher layers; nonetheless, from the
point of view of 1D model atmosphere  structure and  modelling, this
difference is not important. Therefore, from the point  of view of 1D
analysis,  any systematic  effect introduced by an incorrect temperature
scale, by inhomogeneities etc, should affect the Mg II
lines  and  the O  I triplet in a similar way.  Since no systematic
differences were found between Mg I and Mg II abundances, we might say that
the oxygen abundances derived from  the triplet should also be free of
these  atmospheric  problems. In other words, O I triplet-based abundances in metal-poor stars 
may be wrong, but our results concerning Mg II abundances argue against this reasoning.  
This  conclusion must be
confirmed by deriving  magnesium abundances from the Mg  II lines in a
large sample of very metal-poor stars ([Fe/H]$\la -1.5$), where the
oxygen  problem  really exists. Of course, in  3D  model  atmosphere
analyses the  concept of the formation depth of a given line is meaningless
since in some parts of the atmosphere (e.g. warm up-flows) the
line  is  formed  at a given depth while in other parts (e.g. cool
down-flows) the line may be  formed at quite a different depth. However,
for lines having very similar excitation energies and belonging  
to majority  species (as do the Mg II and O I
triplet lines), large
differences in the 3D effects are not expected (M. Asplund, private communication).
Thus, the  conclusion above in 1D may hold when 3D
atmospheres are used. Studies of 3D N-LTE formation of O I and  Mg II lines in
metal-poor stars are urgently needed to confirm this.

\begin{table}
\centering
\caption{Averaged formation depths over the line profile for 
the Mg II lines and O I triplet.}
\begin{tabular}{@{}lcccc@{}}
\hline
\multicolumn{5}{c}{log $\tau_{5000}$}\\
           &           &           &         & \\     
Star/Model & MgII 7877 & MgII 7896 & OI 7774 & OI 7775 \\
\hline
HD 51530   & $-0.20$   & $-0.20$   & $-0.44$ & $-0.38$        \\
HD 59984   & $-0.18$   & $-0.20$   & $-0.40$ & $-0.32$        \\
BD +18$^{\circ}$ 3423& $-0.08$& $-0.12$&$-0.22$& $-0.17$\\
6000/4.0/-2.0& $+0.00$    & $-0.05$   & $-0.14$ & $-0.14$      \\
\hline
\end{tabular}
\end{table}

\begin{figure}
\includegraphics[width=85mm]{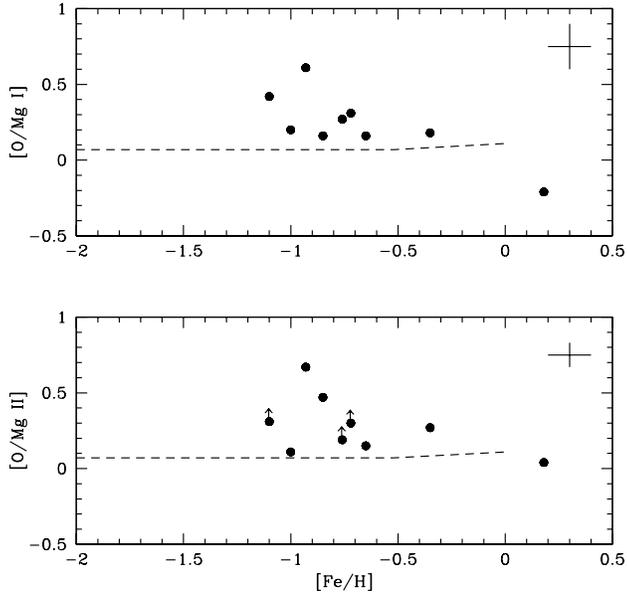}
\caption{[O/Mg] vs. [Fe/H] relationship in the sample stars obtained from N-LTE  
Mg I and Mg II abundances. The dashed line connects the 
predicted [O/Mg] ratio according to the yields  
obtained by Limongi \& Chieffi (2002) from type II supernova models and their 
dependence on the metallicity of the progenitor (see text). Note the different error bar
in [O/Mg]. This is because uncertainties in the atmosphere parameters affect the Mg I and Mg II lines in opposite
senses.}
\end{figure}

As an additional test, we  compared the [O/Mg] vs. [Fe/H] ratios derived
using Mg I and Mg II lines. This is shown in Figure 6.  From the literature, we
took the equivalent width measurements of the O
I  triplet  in stars  in  common with  this  work  and derived  oxygen
abundances  according  to  the stellar parameters shown in  Table  2.  The
spectroscopic data for the O I  triplet were taken from Kurucz \& Bell
(1995). The standard van der Waals damping constant from Uns\"old was
modified to fit the equivalent widths of the O I triplet in the solar
flux spectrum (Kurucz  et al. 1984), adopting a solar oxygen abundance
of  log $\epsilon$(O)$=8.69$ (Allende Prieto et al.  2001). Oxygen abundances  were corrected from  N-LTE  effects
according to the formula given by Takeda (2003) (see his equation 1)
which  takes  into  account  the  dependence of  this correction on
T$_{\rm{eff}}$ and log  $g$\footnote{Takeda  (2003) does not find  any
dependence  of N-LTE corrections on the O  I triplet with the
metallicity.}.  Therefore, Figure 6 shows N-LTE [O/Mg] ratios.

It can be seen that the of  the [O/Mg] ratios from Mg I and Mg II
abundances are positive.  [O/Mg] from Mg II lines are a little higher on
average due to the sign of the N-LTE corrections (see  Sect. 3). The
point at [O/Mg]$\sim  +0.7$ is the star BD  $+18^\circ$ 3423 which has
an unusually low  Mg abundance (see Table 5) considering  that it is a halo
star.  Mashonkina  et  al.  (2003)  also derive a low Mg abundance
([Mg/Fe]$=+0.12$) for this star as well as for other halo stars,
suggesting that the  halo  stellar population  may  be inhomogeneous  by
origin.  In any case, despite the low number of stars and considering the  
errors in the [O/Mg] ratios (see Figure 6), we can say that the [O/Mg]
ratio is  constant in the metallicity  range of our  stars. Carretta et
al. (2000), using both [OI] and O  I lines (but only Mg I lines), find
[O/Mg]  ratios at  the  same  level of  enhancement  and with  similar
scatter  at a given  metallicity in  stars with  $-2.5\la$[Fe/H]$\la
-1$. 

In Figure  6, the dashed line represents  the predicted [O/Mg]
ratio according to the SNII explosion models by Limongi \& Chieffi
(2002). These  SNII models cover the mass $13\la$  M/M$_{\sun}\la 40$
and  metallicity $Z=0.02$  to  10$^{-6}$ ranges. The line was
computed by integrating the yields in the  mass and metallicity ranges
with Salpeter's standard IMF. Clearly, most of  the observed [O/Mg]
values  are well above this line, even considering the error
bars\footnote{Note that using a  higher oxygen  abundance in the Sun
(e.g. 8.87, Grevesse \& Sauval 1998) to derive  the [O/Mg] ratio, the
resulting line  would be $\sim$0.1 dex below.}.  There are at
least two possible explanations of this discrepancy between theory and
observations:

i)  Since most of the [O/Mg]  ratios in the literature
(including the  present work) are derived from the O  I triplet, this
would indeed indicate that the  oxygen abundances derived from these
lines are too large, for  as yet unknown reasons (see Fullbright \& Johnson 2003, for a 
discussion of several alternative explanations). However, note that in
the  metallicity range studied  here, all  the oxygen  indicators agree
using 1D and/or  3D atmosphere models (see Nissen  et al. 2002). Thus,
eventually [O/Mg] ratios derived from oxygen indicators other than the
oxygen  triplet would give similar  ratios  (at least in stars with
$-1.0\la$[Fe/H]$\la 0.0$). However, very recently Bensby et al. (2003) studied
the [O/Mg] vs. [Fe/H] relationship in a sample of stars with metallicity
from [Fe/H]$\approx -0.9$ to $+0.4$ using oxygen abundances from the [OI]
6300 {\AA} line. Their [O/Mg] are in closer agreement with the theoretical
expectations, suggesting indeed that the OI triplet lines are not good
abundance indicators. They noted also that oxygen does not follow
magnesium at super-solar metallicities as the [O/Mg] ratio in the
star with [Fe/H]$>0.0$ in Figure 6 may suggest. 

ii) Oxygen  and/or magnesium yields in SNII
models are  wrong. However, oxygen  and magnesium are  produced mainly
during hydrostatic burning in the SNII progenitor and only a small
fraction of  the ejecta stems from explosive  neon- and carbon-burning
(e.g. Thielemann et  al. 1996). Therefore, their yields  should not be
greatly affected  by uncertainties in the SNII  explosion model. Instead,
the actual value of the $^{12}$C($\alpha,\gamma)^{16}$O reaction rate
and the treatment of rotation and convection during the hydrostatic
phases may affect the O and Mg  yields. This uncertainty, however,
would be translated into a variation of no more than $\pm  0.1$ dex in
the position of the dashed line in Figure 6 (see the discussions by
Argast et al.  2002; Heger \& Langer 2000; Imbriani  et al. 2001), and
the   discrepancy  between theory and observations would remain. 

iii) Finally, we should take into account 
that the formation of spectral lines
in metal-poor atmospheres is not fully understood yet and, therefore, any
conclusion extracted from the [X/Fe] vs. [Fe/H] studies on the chemical
evolution of the Galaxy should be considered with caution. Let us remember
the recent discrepancy found between the observed and the predicted
primordial Li abundances from Big-Bang nucleosynthesis as inferred by the
WMAP analysis of the cosmic microwave background (Bennett et al. 2003).   
Obviously, homogeneous and 3D studies of the [O/Mg] ratio  in a larger sample of stars are
needed to answer this question.

\section{Summary}

We have analysed several Mg I and two Mg II lines in a sample of 19
mildly metal-poor dwarf stars in a search for differences
between these two Mg abundance indicators. We found a good agreement between both in 
the LTE and N-LTE cases. Similar to the features found in other studies, the [Mg/Fe] 
ratios derived here show significant scatter at [Fe/H]$\sim -0.6$ which is currently
ascribed to the existence of thick and thin disk stars at about this metallicity, stars
which are chemically discrete from each other. Using the oxygen abundances from the
literature, we also studied the [O/Mg] vs. [Fe/H] relationship. The [O/Mg]
ratios derived are constant over the range of metallicity studied but significantly larger
than the nucleosynthetic predictions obtained from SNII models. There is no easy explanation
to this problem. It could be both spectroscopic and theoretical but, in any
case, the determination of any abundance ratio in metal-poor stars urgently
requires a detailed modelling (3D) of their atmospheres as well as a simultaneous N-LTE
study. Only thus could we safely draw conclusions concerning
the chemical evolution of the Galaxy.

\section*{Acknowledgements}
Data from the VALD  database at Vienna were used for
the preparation of  this paper. The 2.5m NOT  telescope is operated on
the island of La Palma by the RGO in the Spanish  Observatory of the
Roque  de  los  Muchachos  of the Instituto de  Astrof\'\i  sica  de
Canarias. This work was also based in part on observations collected
with  the 2.2m  telescope at the German-Spanish Astronomical Centre,
Calar Alto (Almer\'\i  a).  It was partially supported  by the Spanish
grant   AYA2002-04094-C03-03  from   the  Ministerio de Ciencia y
Tecnolog\'\i a. ML thanks RBRF (grant  02-02-17174) and the RF President on Leading
Scientific Schools (grant 1789.2003.2) for partial financial support.
We thank the referee for the valuable comments and criticism which
served to improve this work considerably.

\end{document}